# Theoretical investigation, simulation and empirical analysis of the growth pattern of traffic oscillations in the Euler coordinates

Junfang Tian[a], Martin Treiber[b], Rui Jiang[c], Xiaopeng Li[d], Bin Jia[c], Ziyou Gao[c]

[a]Institute of Systems Engineering, College of Management and Economics, Tianjin University, Tianjin 300072, China, Email: jftian@tju.edu.cn

[b]Institute for Transport & Economics, Technische Universität Dresden, Würzburger Str. 35, D-01062 Dresden, Germany, Email: martin@mtreiber.de

[c]MOE Key Laboratory for Urban Transportation Complex Systems Theory and Technology, Beijing Jiaotong University, Beijing 100044, China, Email: jiangrui@bjtu.edu.cn (RJ)

[d]Department of Civil and Environmental Engineering, University of South Florida, FL 33620, USA, Email: xiaopengli@usf.edu

**Abstract**

The formation and development of oscillations is an important traffic flow phenomenon. Recent studies found that along a vehicle platoon described in the Lagrangian specification, traffic oscillations grow in a concave way. Since stationary bottlenecks are more intuitively described in the Eulerian framework, this paper investigates whether the concave growth pattern of traffic oscillations in the Lagrangian coordinates can be transferred to the Euler coordinates (i.e. the concave increase in standard deviation is no longer measured across the vehicle indices but as a function of the road location). To this end, we theoretically unify these two ways of measuring oscillations by revealing their mapping relationship. We show that the growth pattern measured in the Lagrangian coordinates can be transferred to the Euler coordinates. We believe this finding is nontrivial since the scenarios are significantly different: while in vehicle platoons (Lagrangian view, non-penetrable moving bottleneck), the speed variance for a given vehicle is ideally constant, the drivers in the Eulerian setting (penetrable stationary bottleneck triggering the waves) experience all amplitudes, first the big ones and then the small ones. To test this proposition, we performed simulation using two different kinds of car-following models. Simulation results validate the theoretical analysis. Finally, we performed empirical analysis using the NGSIM data, which also validates the theoretical analysis.

## 1. Introduction

Driving in smooth traffic flow makes passengers comfortable, saves fuel consumption and emissions and mitigates rear crash risks. Unfortunately, traffic flow is usually not smooth. Traffic oscillations have been observed frequently, in particular in congested traffic flow (see e.g., Kerner, 1998, 2002, 2009, 2017; Yeo and Skabardonis; 2009; Mauch and Cassidy, 2002; Schönhof and Helbing, 2007; Yuan et al. 2017; Arnesen and Hjelkrem, 2018). A recent Japanese experiment (Sugiyama et al., 2008) verified that when the density exceeds a threshold, traffic flow is always unstable. Traffic oscillations emerge and finally develop into traffic jam.

For over half a century, it is believed that the linear instability of traffic steady state leads to the growth of oscillations (see e.g., Treiber and Kesting, 2013; Wilson and Ward, 2011; Nagel et al. 2003; Nagatani, 2002; Helbing, 2001). Li et al. (2012, 2014) and Li and Ouyang (2011) showed with both theoretical and empirical results that nonlinear driving behavior may contribute to concave growth of traffic oscillations at least after oscillation amplitude becomes significant. Jiang et al. (2017, 2015, 2014) performed several car-following experiments on the open road section, in which the leading car moves with different constant speed to simulate moving bottleneck. They observed the formation and development of oscillations, and found that along the platoon, standard deviations (STDs) of the speed increase concavely. Later, to investigate the evolution of traffic oscillations, Tian et al. (2016) analyzed the NGSIM vehicle trajectories data on US 101 highway. They found that traffic oscillations measured from empirical data also grow concavely highly consistent with the car following model predictions.

Simulation was performed (Jiang et al., 2015, 2014) using traditional car-following models, including the Intelligent Driver Model (IDM, Treiber et al., 2000), the Full Velocity Difference Model (FVDM, Jiang et al., 2001), the Gipps Model (Gipps, 1981), the Optimal Velocity Model (Bando et al., 1995), and the General Motor models (Chandler et al., 1958; Edie, 1961; Gazis et al. 1961). A fundamental assumption in these models are: under steady state conditions, there is a one-to-one relationship between vehicle speed and spacing. Simulations show that in the unstable condition in these models, the STD initially increases in a convex way, which runs against the experimental finding. It was shown that if the fundamental assumption is removed and the traffic state is allowed to span a two-dimensional region in speed-spacing plane, the growth pattern of oscillations would change and become consistent with the experimental finding qualitatively or even quantitatively. In particular, the two-dimensional Intelligent Driver Model (2D-IDM), in which the desired time gap changes with time, can quantitatively reproduce the experimental results.

The finding that traffic oscillations grow concavely is important, which suggests that the instability mechanism of traffic flow in traditional models is questionable. In traditional models, the traffic flow becomes unstable because the steady state solution is linearly unstable. In contrast, based on the experimental results, Jiang et al. (2017) argued that the traffic flow instability comes from the cumulative effect of stochastic factors, see also Treiber and Kesting (2017), Laval et al. (2014). As Li et al. (2014) proved, the linear instability in a nonlinear Newell model results in initial convex growth of oscillations. It has been further demonstrated that in the traditional models assuming a one-to-one relationship between spacing and speed, oscillations increase convexly in the initial stage (Tian et al., 2016), which was not consistent with the observed concave growth pattern.

Moving bottlenecks are more intuitively described in the Lagrangian specification co-moving with the bottleneck while stationary bottlenecks are better described in the Eulerian specification. Naturally, the question arises whether the concave growth pattern within a vehicle platoon in the Lagrangian coordinates can be transferred to the settings of a fixed bottleneck in the Euler coordinates where the concave increase in speed STD is no longer measured across the vehicle indices but along the road longitudinal location.

To address this issue, this paper reveals the relation between the propagation patterns of traffic oscillations upstream of moving and stationary bottlenecks. Firstly, we theoretically unify these two measures of oscillations in different coordinate systems by revealing their mapping relationship and demonstrate that the growth pattern measured in the Lagrangian coordinates can be transferred to the Euler coordinates. Next, we performed simulation using two different kinds of car-following models. The simulation results validate the theoretical analysis. Finally, we performed an empirical study on the growth pattern of oscillations in traffic flow on US 101 freeway, which showed that oscillations do grow concavely along the road. This further validates the theoretical analysis.

The paper is organized as follows. The next section theoretically investigates the relation of oscillation growth pattern between Euler and Lagrangian coordinates. Section 3 performs simulation studies of the oscillation growth pattern along the road. Section 4 analyzes the oscillation growth pattern along the road in empirical data, and Section 5 concludes the paper.

**2. Theoretical investigation**

The investigated oscillation growth in the Euler coordinates can be essentially related to that in the Lagrangian coordinates characterized in earlier studies (Tian, et al. 2016). This section will theoretically unify these two measures of oscillations in different coordinate systems by revealing their mapping relationship. Specifically, we will compare temporal averages along single trajectories (Lagrangian view) with spatiotemporal averages considering all trajectories in a rectangular spatiotemporal region centered around a certain location (Eulerian view).

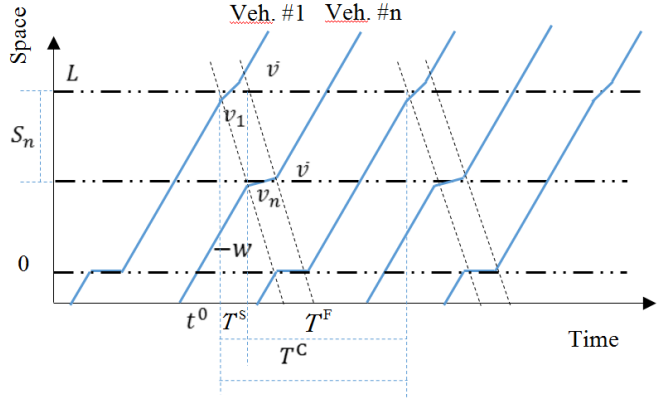

Fig. 1. Schematic model of oscillations along trajectories.

Figure 1 presents a schematic model of oscillations along trajectories that approximates the basic periodic patterns. Basically, we assume that along the time horizon, trajectories repeat oscillation patterns with period $T^C$. Consider a highway segment between a downstream location $L$ (near a bottleneck where the oscillation initiates) and an upstream location 0 and a typical period starting at $t^0$. We index the lead vehicle that initiates oscillation during this period with Vehicle 1, and increase the vehicle index from downstream to upstream. Now, we assume that the oscillation propagates backward along a shockwave at a constant velocity $-w$ and the upstream vehicles follow a similar pattern with gradually amplified oscillations. We investigate the spatiotemporal region of trajectories between the shock wave that starts this period and the one that ends it. We call the region of slow trajectory segments the slow band and the region of cruising trajectories the cruise band. Along the propagation direction, the slow-moving speed $v_n$ for Vehicle $n$ in the slow band decreases with the vehicle index, while the cruising speed $\hat{v}$ in the cruise band is independent of the vehicle index. Let $s_n$ denote the average spacing between vehicles $n-1$ and $n$(including the vehicle length) during the slow phase of the oscillation. Then, with simple geometry, we can calculate the distance of vehicle $n$'s position from location $L$ at the beginning of the slow phase, denoted by $S_n$, as follows:

$$S_n := \sum_{n'=2}^{n} \frac{w s_{n'}}{v_{n'-1}+w}. \tag{1}$$

Furthermore, the slow band reaches Vehicle $n$ at time $t_n^0$, which is given by

$$t_n^0 := t^0 + \frac{S_n}{w} = t^0 + \sum_{n'=2}^{n} \frac{s_{n'}}{v_{n'-1}+w}. \tag{2}$$

Note that the Euler speed distribution at location $L - S_n$ during period $[t_n^0, t_n^0 + T^C]$ is related to that of the Lagrangian speed measurements of vehicle $n$ during this trajectory band in the following way. While the Euler measurements at location $L - S_n$ stay at a speed close to $v_n$ during $[t_n^0, t_n^0 + T^S]$ and stay at $\acute{v}$ during $[t_n^0 + T^S, t_n^0 + T^C]$, the Lagrangian speed measurements of vehicle $n$ shall be at $v_n$ during time $[t_n^0, t_n^0 + t_n^S]$ and stay at $\acute{v}$ during $[t_n^0 + t_n^S, t_n^0 + t_n^C]$ where

$$t_n^C := t_n^S + t_n^F, \; t_n^S := \frac{T^S w}{w+v_n}, \; t_n^F := \frac{T^F w}{w+\acute{v}} \tag{3}$$

This yields vehicle $n$'s average speed

$$\bar{v}_n^L := \frac{v_n t_n^S}{t_n^C} + \frac{\acute{v} t_n^F}{t_n^C},$$

and its Lagrangian speed STD is

$$\sigma_n^L := \sqrt{(v_n - \bar{v}_n^L)^2 \, t_n^S/t_n^C + (\acute{v} - \bar{v}_n^L)^2 \, t_n^F/t_n^C}. \tag{4}$$

This can be interpreted as a weighted operation of speeds $v_n$ and $\acute{v}$ with weights $t_n^S/t_n^C$ and $t_n^F/t_n^C$, respectively.

In the Euler case, we examine the speed STD calculated from all trajectories in a fixed region (small "lattice cell") $S_n$ during an oscillation cycle $[t_n^0, t_n^0 + T^C]$. Based on the conservation law, the vehicle density in the slow moving band $[t_n^0, t_n^0 + T^S]$ (where vehicles cross location $S_n$ at speed close to $v_n$), denoted by $\rho_n^S$, and the vehicle density in the cruise band $[t_n^0 + T^S, t_n^0 + T^C]$, denoted by $\rho_n^F$, satisfy the following simple relationship:

$$\frac{\rho_n^S}{\rho_n^F} = \frac{w+\acute{v}}{w+v_n}. \tag{5}$$

Since the average vehicle number in the lattice cell is proportional to the density, the free and congested states are not only weighted with the corresponding Eulerian time intervals but also with the densities. Therefore, the average speed is given by

$$\bar{v}_{S_n}^E := v_n \frac{T^S \rho_n^S}{T^S \rho_n^S + T^F \rho_n^F} + \acute{v} \frac{T^F \rho_n^F}{T^S \rho_n^S + T^F \rho_n^F}.$$

With this, similar to Equation (4), the Euler speed STD detected at location $S_n$, denoted by $\sigma_n^E$, shall be a weighted operation of speeds $v_n$ and $\acute{v}$ with weights $\frac{T^S \rho_n^S}{T^S \rho_n^S + T^F \rho_n^F}$ and $\frac{T^F \rho_n^F}{T^S \rho_n^S + T^F \rho_n^F}$, respectively:

$$\sigma_{S_n}^E := \sqrt{\left(v_n - \bar{v}_{S_n}^E\right)^2 \frac{T^S \rho_n^S}{T^S \rho_n^S + T^F \rho_n^F} + \left(\acute{v} - \bar{v}_{S_n}^E\right)^2 \frac{T^F \rho_n^F}{T^S \rho_n^S + T^F \rho_n^F}}. \tag{6}$$

Note that with Equations (3) and (5), it is easy to see that $\frac{T^S \rho_n^S}{T^S \rho_n^S + T^F \rho_n^F} = t_n^S/t_n^C$ and $\frac{T^F \rho_n^F}{T^S \rho_n^S + T^F \rho_n^F} = t_n^F/t_n^C$. Remarkably, this shows that the Euler speed STD are equivalent to the Lagrangian speed STD measures along individual vehicle trajectories:

$$\bar{v}_n^L = \bar{v}_{S_n}^E, \sigma_n^L = \sigma_{S_n}^E. \tag{7}$$

With this, if we are given Langrangian STDs $\sigma_n^L$ where $\sigma_n^L$ is the STD for Vehicle $n$, then the corresponding Euler STDs $\sigma_S^E$ where $\sigma_S^E$ is the STD at location $L-S$ can be estimated in the following way. First use Equation (1) to obtain $n$ such that $S \in [S_{n-1}, S_n]$. Then we can simply obtain $\sigma_S^E$ as an

8interpolation of $\sigma_n^L$ and $\sigma_{n-1}^L$. On the other hand, if we are given Euler STDs $\{\sigma_{S^i}^E\}$, then the corresponding Langrangian STDs $\sigma_n^L$ can be estimated in the similar way. Given a vehicle $n$, again find the corresponding distances $S^{i-1}$ and $S^i$ such that $S_n \in [S^{i-1}, S^i]$. Then we can interpolate $\sigma_{S^i}^E$ and $\sigma_{S^{i-1}}^E$ to obtain the $\sigma_n^L$. With this simplified relationship, we can see that the concavity growth of Lagrangian data implies that of Euler data, and vice versa.

We would like to emphasize that the concavity survives the transition from Lagrangian to Eulerian situations is nontrivial. The scenarios are significantly different: while in vehicle platoons (Lagrangian view, non-penetrable moving bottleneck), the speed variance for a given vehicle is ideally constant, the drivers in the Eulerian setting (penetrable stationary bottleneck triggering the waves) experience all amplitudes, first the big ones and then the small ones.

## 3. Simulations

This section carries out simulations using two typical types of models. The IDM and the FVDM are two representative traditional models, in which the oscillation exhibits a S-shape growth pattern along the platoon. It grows initially in a convex way. When the oscillation amplitude grows to certain extend, the convex growth transits into concave growth. In the 2D-IDM, the oscillation grows concavely along the platoon. In the Stochastic IDM (SIDM), the oscillation exhibits a S-shape growth pattern as in traditional models when the noise is weak. However, when the noise amplitude is large enough, the oscillation grows concavely (Treiber and Kesting, 2017).

To mimic the bottleneck in the NGSIM US 101 dataset (Chen et al., 2012), a rubbernecking effect is implemented to examine the growth pattern of traffic oscillations along the road. In the simulations, the road length is set to 2200 m and the rubbernecking location is at 1800 m. When a vehicle passes the location, the driver rubbernecks with probability *γ*. If a driver rubbernecks, the vehicle speed decreases with deceleration $d_r$ for $T_d$ seconds.

The leading vehicle in the road section will be removed when its location $x_{leading} > 2200$ m. Its following vehicle becomes the new leading one and moves freely. The inflow condition is set as follows. Denote the location and speed of the last vehicle as $x_{last}$ and $v_{last}$, respectively. When $x_{last} > x_0$, a new vehicle will be inserted into the road at $x = 0$. $x_0$ and the speed of the new vehicle $v_{new}$ will be specified in the following subsections.

To investigate the STDs of speed along the road, the study area (0 to 1800 m) is classified into 90 lattices of length 20 m, with the entrance of the road segment being the first lattice and the rubbernecking bottleneck being the last lattice. The speed STDs along all trajectories in each lattice are calculated.

## 3.1 The IDM

In the IDM (Treiber et al., 2000), the acceleration $a_n$ of vehicle $n$ is calculated as follows

$$a_n(t) = a_{max}\left(1 - \left(\frac{v_n(t)}{v_{max}}\right)^4 - \left(\frac{d_{n,desired}(t)}{d_n(t)}\right)^2\right) \tag{8}$$

$$d_{n,desired}(t) = \max\left(v_n(t)T - \frac{v_n(t)\Delta v_n(t)}{2\sqrt{a_{max}b}}, 0\right) + d_0 \tag{9}$$

here $v_n$ is vehicle $n$'s speed; $d_n(t) = x_{n+1}(t) - x_n(t) - L_{veh}$ is its spacing to the leading vehicle $n+1$, $L_{veh}$ is the length of the vehicles, $x_n$ and $x_{n+1}$ are locations of vehicle $n$ and $n+1$, respectively; $\Delta v_n(t) = v_{n+1}(t) - v_n(t)$ is the relative speed; $v_{max}$ is the maximum speed of vehicles, $a_{max}$ and $b$ are the maximum acceleration and the comfortable deceleration, respectively, $T$ is the desired time gap, and $d_0$ is the minimum gap in jams.

Figure 2(a) shows the formation and evolution of oscillations induced by the bottleneck. In this case, $\gamma = 0.95$. The bottleneck is strong, and the capacity is 470 vehicles/h. Figure 2(b) shows the speed STDs at the location from 0 to 1800 m. The S-shape growth pattern is observed along the road.

Figure 2(c) shows the situation induced by a weaker bottleneck with $\gamma = 0.7$, corresponding to higher bottleneck capacity 710 vehicles/h. In this case, the mean speed in the lattice of the rubbernecking region is 10 km/h. Simulation indicates that when the leading car of a platoon moves with this speed, speed STD corresponding to the convex-concave transition point is about 0.95 m/s, see Fig.3(a). The speed STD at the first virtual detector is already 1.14 m/s, which exceeds that of the convex-concave transition point. As a result, only the concave growth pattern has been observed, see Fig.2(d).

The simulation thus validates the theoretical analysis that the concave/convex growth measured in the Lagrangian coordinates can be transferred to the Euler coordinates.

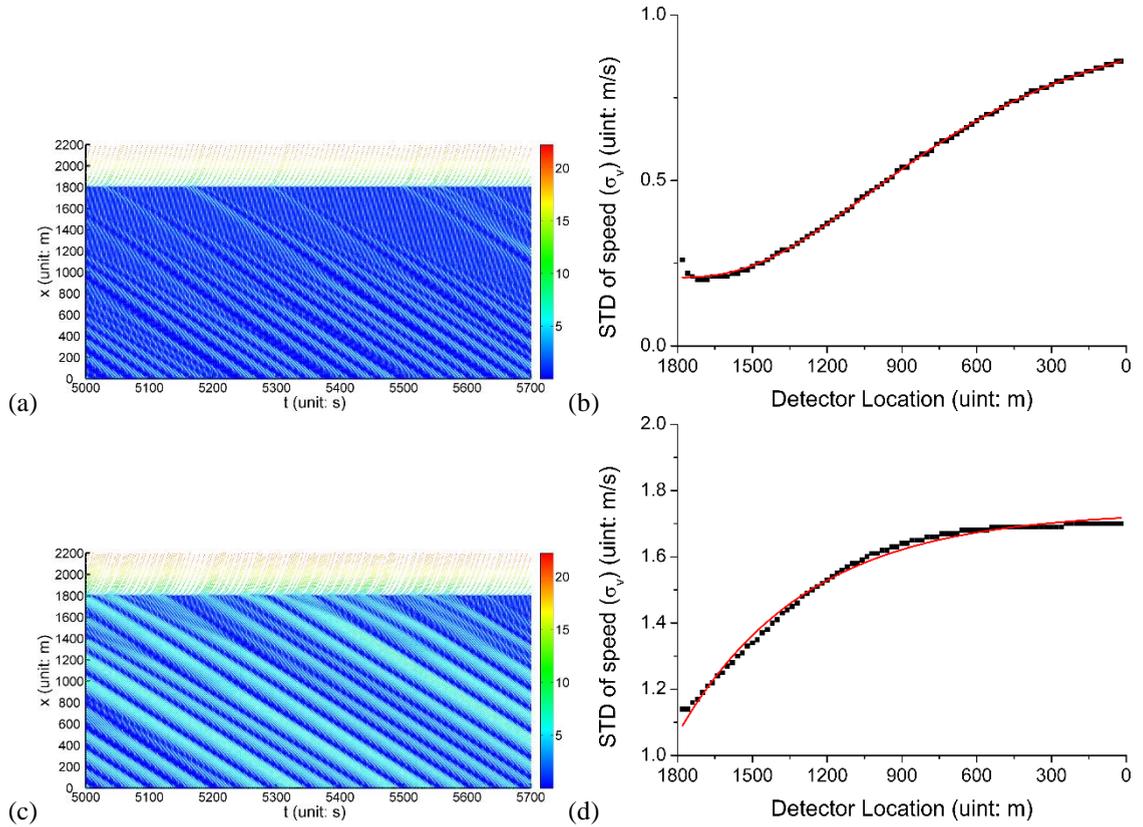

Fig. 2. Simulation results of the IDM. The parameters are set to: $a_{max}= 0.73$ m/s$^2$, $b=1.67$ m/s$^2$, $T =1.6$ s, $L_{veh} = 5$ m, $d_0 =2.0$ m, $v_{max}= 80$ km/h, $d_r = 1$ m/s$^2$, $T_d = 2.5$ s, $x_0 = v_{last}T$, $v_{new} = v_{last}$. (a,c) The speed (unit: m/s) spatiotemporal diagram. (b,d) The STD of speed at each detector, in which the red line is the fitted curve. In (a,b), $\gamma =0.95$; in (c,d), $\gamma = 0.5$.

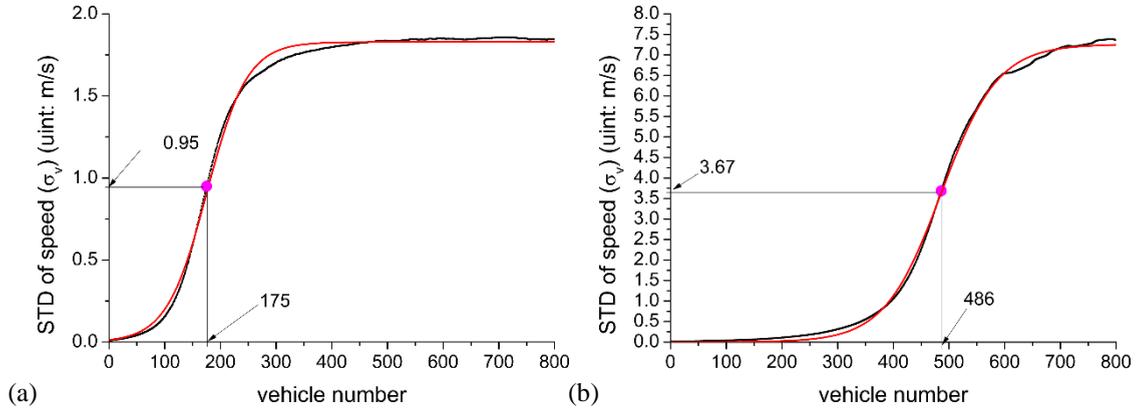

Fig. 3. The speed STDs of vehicles along a vehicle platoon. (a) the simulation results of the IDM and the speed of the leading vehicle is 10 km/h. (b) the simulation results of the FVDM and the speed of the leading vehicle is 60 km/h. The red dots are the transition points from convex to concave.

*3.2 The FVDM*

In the FVDM (Jiang et al., 2001), the acceleration $a_n$ of vehicle $n$ is calculated as follows

$$a_n(t) = \kappa \left( V_{opt}\left(x_{n+1} - x_n\right) - v_n(t) \right) + \lambda \Delta v_n(t) \tag{10}$$

here $v_n$ is vehicle $n$'s speed; $x_n$ and $x_{n+1}$ are locations of vehicle $n$ and its preceding one $n+1$, respectively; $\Delta v_n(t) = v_{n+1}(t) - v_n(t)$ is the relative speed; $\kappa$ and $\lambda$ are two sensitivity parameters; $V_{opt}$ is optimal velocity function and is set to

$$V_{opt}\left(x_{n+1} - x_n\right) = 11.6 \left( \tanh\left(0.086\left(x_{n+1}(t) - x_n(t) - 25\right)\right) + 0.913 \right) \tag{11}$$

as in Jiang et al. (2014).

Fig.4 shows the simulation results, which are similar to that of IDM. The S-shape growth pattern is observed when the bottleneck is strong and the initial speed STD is small, see Fig.3(b); When bottleneck becomes weaker, the mean speed in the lattice of the rubbernecking region is 60 km/h, and the initial speed STD equals 3.92 m/s, exceeding that of the convex-concave transition point (which is about 3.67 m/s based on platoon simulation, see Fig.3(b)). Thus, only concave growth pattern is observed, see Fig.3(d). These again validate the theoretical analysis.

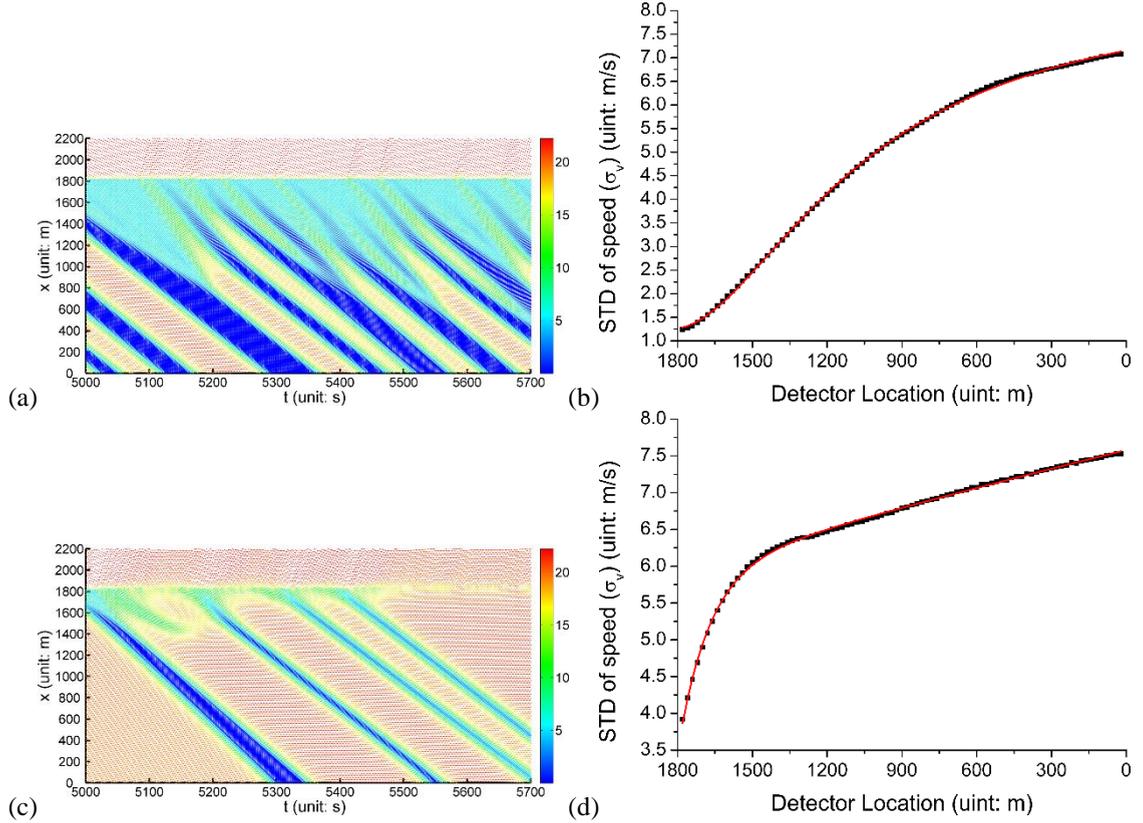

Fig. 4. Simulation results of the FVDM. The parameters are set to: $\kappa = 0.2$ s$^{-1}$, $\lambda = 0.5$ s$^{-1}$, $d_r = 1$ m/s$^2$, $T_d = 2.5$ s, $x_0 = d$, $v_{\text{new}} = V_{opt}(d)$, $d = 30$ m. (a, c) The speed (unit: m/s) spatiotemporal diagram. (b, d) The STD of speed at each detector, in which the red line is the fitted curve. In (a, b), $\gamma = 0.95$; in (c, d), $\gamma = 0.5$.

*3.3 The stochastic IDM*

In the Stochastic IDM (SIDM), the acceleration noise is added into the acceleration of IDM, which is calculated as follows

$$a_n(t) = a_n^{\text{IDM}}(t) + \xi(t) \tag{12}$$

where $\xi \in [-\sigma, \sigma]$ is a uniformly distributed random number.

Figure 5 shows the simulation results of SIDM. One can see that when the noise is weak, the S-shape growth pattern is observed, see Fig.5(b). In contrast, when the noise is strong, the concave growth pattern is observed, see Fig.5(d). These are consistent with the theoretical analysis.

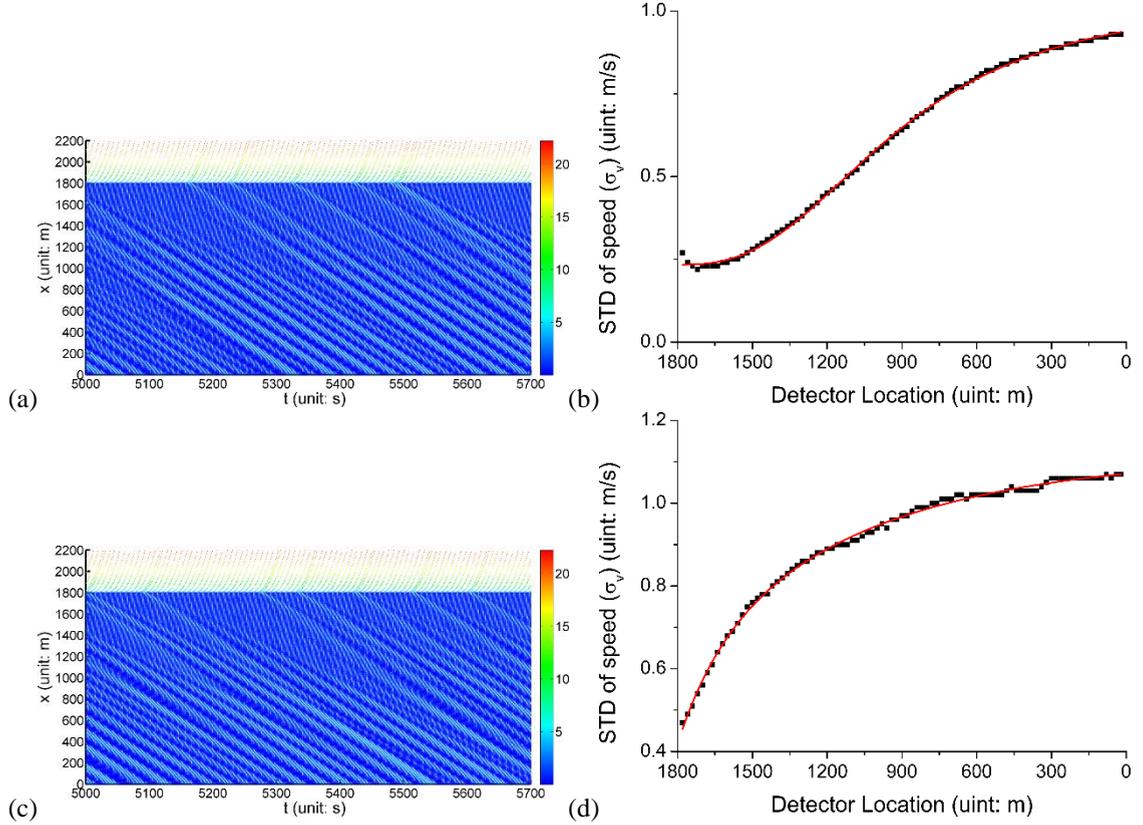

Fig. 5. Simulation results of the SIDM. The parameters are set to: $a_{max}$= 0.73 m/s$^2$, $b$=1.67 m/s$^2$, $T$ =1.6 s, $L_{veh}$ = 5 m, $d_0$ =2.0 m, $v_{max}$=80 km/h, $\gamma$ = 0.95, $d_r$ = 1 m/s$^2$, $T_d$ = 2.5 s, $x_0 = v_{last}T$, $v_{new} = v_{last}$. (a, c) The speed (unit: m/s) spatiotemporal diagram. (b, d) The STD of speed at each detector, in which the red line is the fitted curve. In (a, b), $\sigma$ = 0.2 m/s$^2$; in (c, d) $\sigma$ =1.4 m/s$^2$.

*3.4 The 2D-IDM*

Comparing with the IDM in which the desired time gap *T* is a constant, the 2D-IDM proposed by Jiang et al. (2014) assumes that in each simulation step $\Delta t$ = 0.1 s, the desired time gap *T* changes its value stochastically as follows:

$$T(t+\Delta t) = \begin{cases} T_1 + rT_2 & \text{with probability } p, \\ T(t) & \text{otherwise.} \end{cases} \quad (13)$$

where $r \in [0,1]$ is a uniformly distributed random number, $T_1$ and $T_1 + T_2$ denote the lower and upper threshold of the desired time gap. Variation of the desired time gap enables traffic state to span a two-dimensional region in the speed-spacing plane.

Figure 6 shows the simulation results of 2D-IDM, in which the concave growth pattern is observed along the road. This is also consistent with the theoretical analysis.

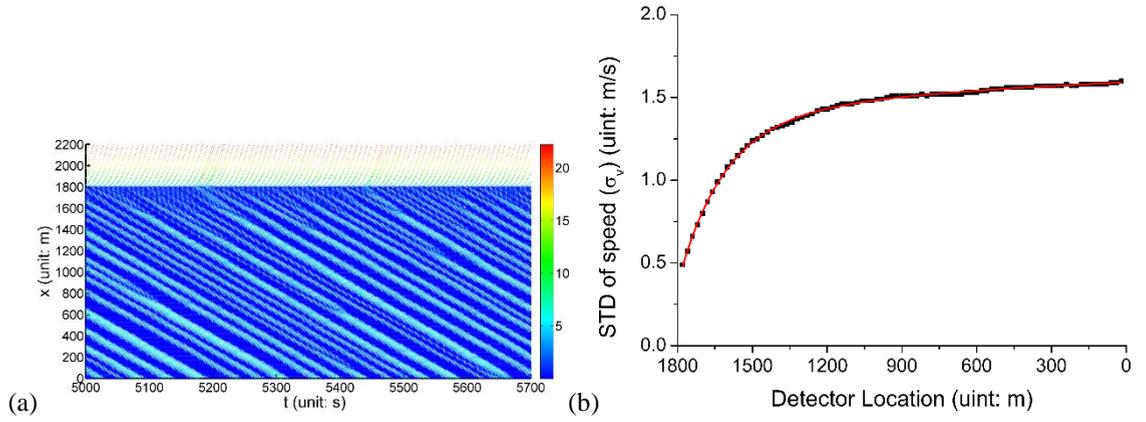

Fig. 6. Simulation results of the 2D-IDM. The parameters are set to: $a_{max}= 0.73$ m/s$^2$, $b =1.67$ m/s$^2$, $T_1 = 0.5$ s, $T_2 = 1.4$ s, $p = 0.015$, $L_{veh} = 5$ m, $d_0 = 2.0$ m, $v_{max}=80$ km/h, $\gamma = 0.95$, $d_r = 1$ m/s$^2$, $T_d = 2.5$ s, $x_0 = v_{last}T_e$, $v_{new} = v_{last}$, $T_e =1.6$ s. (a) The speed (unit: m/s) spatiotemporal diagram. (b) The STD of speed at each detector, in which the red line is the fitted curve.

## 4. Empirical data analysis

We analyze the traffic flow on the south-bound direction of the US-101 Freeway. The trajectory data were collected in the Next Generation Simulation (NGSIM, 2006) project. The data collection segment is 640 m long, near Lankershim Avenue in Los Angeles, California. The data were collected from 07:50 a.m. to 08:35 a.m. on June 15th, 2005. Figure 7 shows a sketch of the data collection location. The main road consists of five lanes, and there is an auxiliary lane that connects to the on-ramp and off-ramp.

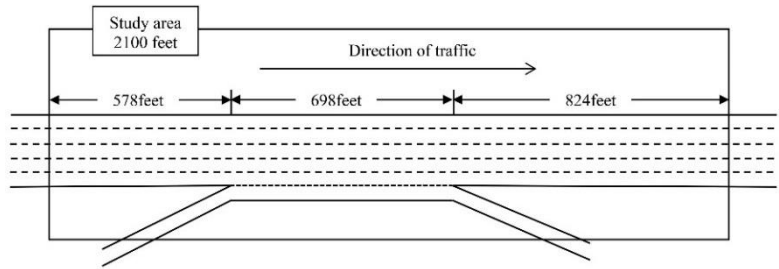

Fig. 7. Sketch of the study area.

We analyze the oscillations on the leftmost lane. To measure the oscillations along the road in a spatiotemporal way, we divide the study area (0 to 600 m) into 30 lattices of length 20 m, with the entrance of the road segment being the first lattice and the exit being the last lattice. As illustrated in Figure 8, we calculate the STDs and the means of the speeds along all trajectories in each lattice during the time interval [0, 770] s. Since the trajectories in the time interval [770, 1000] s involve jams that propagate from the downstream area of the studied section, they are excluded for the analysis. Figure 9(a) shows the means of the speeds. It indicates that the origin of the oscillations should be at the 18th lattice, since the mean speeds downstream of this location show a dramatically increasing trend compared with that upstream of this location. Figure 9(b) shows the STDs from the first lattice to the 18th one. One can see that along the road, the speed STDs grow concavely. The result thus further validates our theoretical analysis.

We would like to mention that in the NGSIM data, the speed STD has reached 3.0 m/s just upstream of the bottleneck. Therefore, the initial growth pattern of traffic oscillation is still unknown. Therefore, in the future work, more empirical and experimental data are needed to investigate the initial growth pattern of oscillations along the road.

One might argue that there exists lane changing in the NGSIM dataset, which would affect the oscillation growth pattern. Actually, simulations on a multilane road indicate that lane changing may trigger oscillation, but it does not have a qualitative effect on the oscillation growth pattern. The simulation results will be reported elsewhere since they are not the scope of this paper.

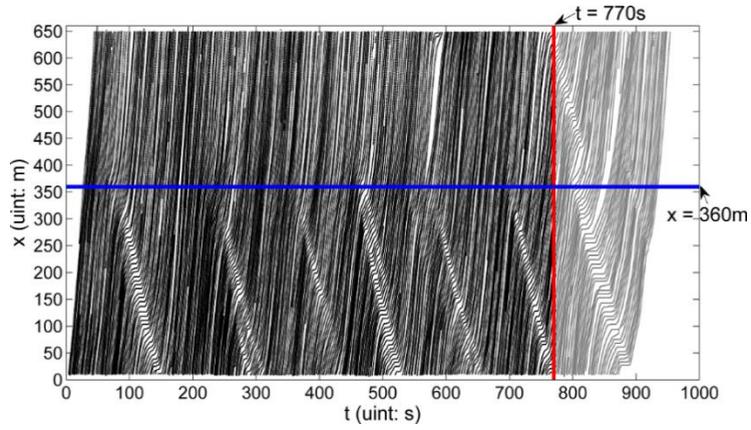

Fig. 8. The NGSIM trajectories on the leftmost lane collected during the time interval 07:50am-08:05am.

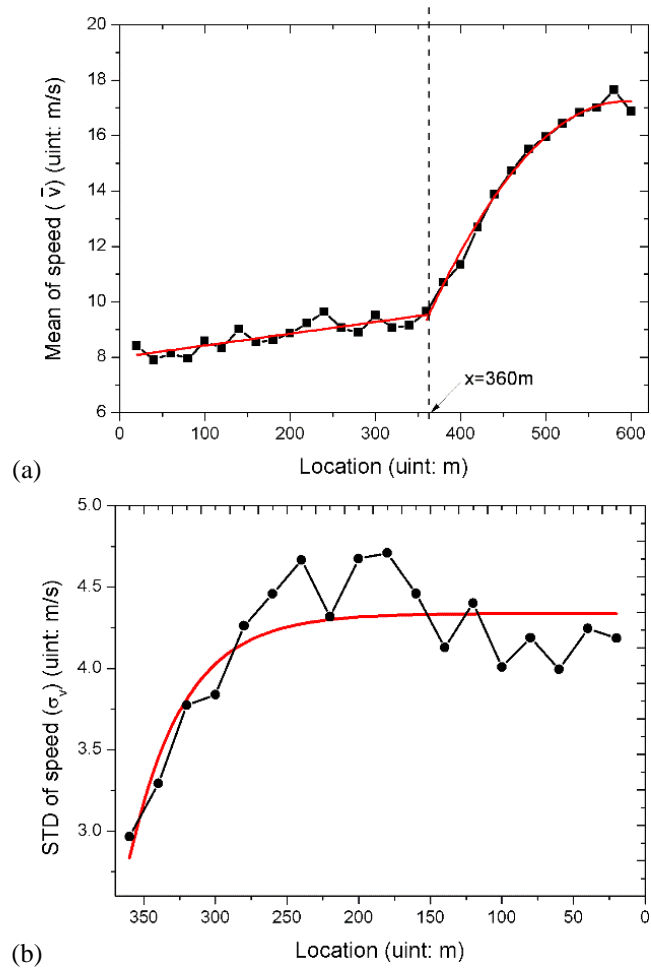

Fig. 9. (a) The mean of speed and (b) the STD of speed at each lattice. The red line is the fitted curve.

## 5. Conclusion

In traffic flow dynamics, propagation and growth of traffic oscillations play an important role. However, due to lack of high-precision trajectory data, the propagation mechanism of traffic oscillations is still unclear. While it is traditionally believed that growth of oscillations is because of the linear instability of traffic steady state, recent efforts (Laval et al., 2014; Treiber and Kesting, 2017; Jiang et al., 2014, 2015, 2017) indicate that stochastic factors lead to the growth of oscillations. This is because experimental and empirical data revealed that along the platoon, oscillations grow concavely, while traditional models reproduce an initial convex growth of oscillations.

In previous studies (Tian et al., 2016; Jiang et al., 2014, 2015, 2017), the growth of oscillations is measured in the Lagrangian view. However, stationary bottlenecks are better described in the Eulerian framework. The theoretical analysis in this paper shows that the concave growth pattern measured in the Lagrangian coordinates can be transferred to the Euler coordinates. Simulations of two difference types of models as well as the empirical study on the oscillation growth in the NGSIM data validates the theoretical analysis. Our study thus sheds further light the evolution mechanism of traffic oscillations.

**Acknowledgements**

This work was supported by the NSFC (Grant Nos. 71771168, 71621001), and the US NSF through Grant CMMI# 1453949. Correspondence and requests for materials should be addressed to RJ and XPL.